\documentclass{article}
\usepackage[preprint]{spconf}
\usepackage{amsmath,graphicx,mathtools,float}
\usepackage{booktabs}
\usepackage{multirow}
\usepackage{xcolor}
\usepackage{color}
\usepackage{algorithm}
\usepackage{algpseudocode}

\copyrightnotice{\copyright\ IEEE 2020}\toappear{To appear in {\it Proc.\ ICASSP2020,
    May 04-08, 2020, Barcelona, Spain}}

\newcommand{\mypar}[1]{{\bf #1.}}

\newcommand{\appropto}{\mathrel{\vcenter{
  \offinterlineskip\halign{\hfil$##$\cr
    \propto\cr\noalign{\kern2pt}\sim\cr\noalign{\kern-2pt}}}}}
    
\title{Damage-sensitive and domain-invariant feature extraction for vehicle-vibration-based bridge health monitoring}
\name{Jingxiao Liu$^{\star}$ \quad Bingqing Chen$^{\dagger}$ \quad Siheng Chen$^{\ddagger}$ \quad Mario Berg\'{e}s$^{\dagger}$ \quad Jacobo Bielak$^{\dagger}$ \quad HaeYoung Noh$^{\star}$}

			\address{$^{\star}$ Civil and Environmental Engineering, Stanford University \\
			$^{\dagger}$ Civil and Environmental Engineering, Carnegie Mellon University \\
			    $^{\ddagger}$ Mitsubishi Electric Research Laboratories (MERL)}
\begin{document}
%
\maketitle
\begin{abstract}
We introduce a physics-guided signal processing approach to extract a damage-sensitive and domain-invariant (DS~\&~DI) feature from acceleration response data of a vehicle traveling over a bridge to assess bridge health. Motivated by indirect sensing methods’ benefits, such as low-cost and low-maintenance, vehicle-vibration-based bridge health monitoring has been studied to efficiently monitor bridges in real-time. Yet applying this approach is challenging because 1) physics-based features extracted manually are generally not damage-sensitive, and 2) features from machine learning techniques are often not applicable to different bridges. Thus, we formulate a vehicle bridge interaction system model and find a physics-guided DS~\&~DI feature, which can be extracted using the synchrosqueezed wavelet transform representing non-stationary signals as intrinsic-mode-type components. We validate the effectiveness of the proposed feature with simulated experiments. Compared to conventional time- and frequency-domain features, our feature provides the best damage quantification and localization results across different bridges in five of six experiments.
\end{abstract}

\section{Introduction}
\label{sec:intro}
\vspace{-2mm}
Bridges are key components of transportation infrastructure, albeit one in eleven bridges in the U.S. were structurally deficient~\cite{asce}. The high cost and time usually required to inspect aging bridges desperately calls for advanced sensing and data analysis techniques.

\begin{figure}[ht]
\begin{minipage}[b]{1\linewidth}
\label{ISHM}
  \centering
  \centerline{\includegraphics[width=8cm]{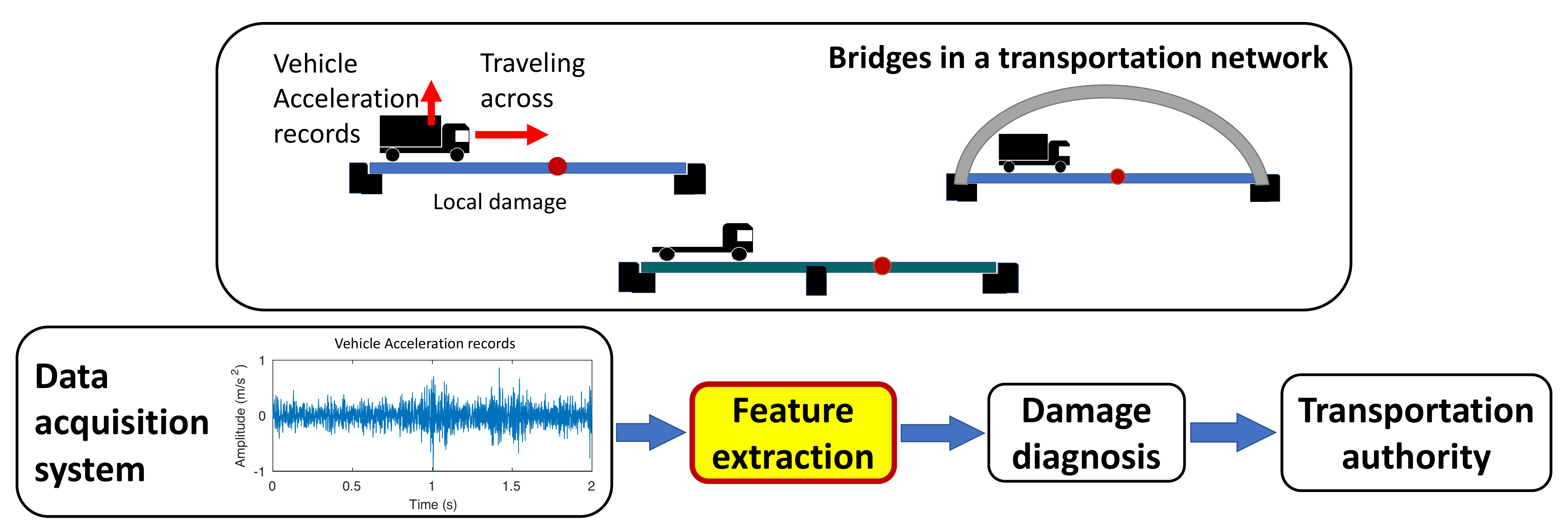}}
\end{minipage}
  \vspace{-7mm}
\caption{Vehicle-vibration-based BHM. Features extracted from vehicle vibration signals are used to diagnose damages. 
}
\label{fig:vbi}
  \vspace{-2mm}
\end{figure}

The use of vibration signals collected from travelling vehicles to monitor structures (Figure~\ref{ISHM}) has recently become a viable alternative to traditional structural health monitoring approaches~\cite{lederman2017track,lederman2017sparse,lederman2017data,liu2019dynamic,liu2019detecting,liu2019damage}.  This approach does not require intensive deployment and maintenance. Previous work on vehicle-vibration-based bridge health monitoring (or indirect BHM) mainly falls into two categories: modal analysis and data-driven approaches. Modal analysis focuses on identifying modal parameters of a bridge, such as natural frequencies ~\cite{yang2005vehicle,yang2015extraction,matarazzo2018scalable}, mode shapes~\cite{yang2014constructing,matarazzo2018scalable} and damping~\cite{yang2019bridge}.  Data-driven approaches use signal processing and machine learning techniques to extract informative features for diagnosing damage~\cite{liu2019damage,lederman2014damage,malekjafarian2019machine}. \vspace{-0.5em}

However, to make the indirect BHM (IBHM) approaches practical, there are three main challenges to address. First, since the vibration signals are indirect measurements of structure's vibrations, modal parameters identified using modal analysis, e.g.,\cite{yang2005vehicle,yang2015extraction,yang2014constructing,matarazzo2018scalable,yang2019bridge}, are sensitive to the vehicle’s properties, environmental factors and noise. Second, purely data-driven methods, e.g., \cite{liu2019damage,lederman2014damage,malekjafarian2019machine,liu2020diagnosis}, can suffer from overfitting to the available data (i.e., a set of bridges with known damage labels) and achieve significantly worse performance when applied to other bridges. Third, labeled data are limited in quantity. Especially for full-scale bridges, it is expensive, time-consuming, and impractical to obtain vehicle vibrations with corresponding damage labels. It is also unrealistic to damage bridges for sourcing damage labels.

We introduce a physics-guided signal-processing algorithm to extract an informative feature, which can estimate and localize damage in a bridge. To handle the first challenge, the extracted feature should be sensitive to damage instead of to uncertainties. In this work, we use the synchrosqueezed wavelet transform (SWT) \cite{daubechies2009synchrosqueezed,daubechies2011synchrosqueezed} to represent the vehicle acceleration signal in the time-frequency plane and reconstruct a DS component using the inverse SWT (ISWT) within a frequency band determined by pre-identified system properties. We use the SWT and ISWT because SWT can represent the non-stationary and time-varying vehicle acceleration as a superposition of intrinsic mode functions (IMF)-type components (our desired feature has the same type), and ISWT can reconstruct our non-stationary component within a frequency range without having the mode mixing effect~\cite{mandic2013empirical}. To addresses the second and the third challenges, a DS~\&~DI feature is obtained by multiplying the reconstructed component by a DI factor. It is obtained from the solution of a vehicle bridge interaction system (VBIS). Because this feature is DI and not extracted by a trainable model with the supervision of damage states, diagnosing damage using this feature does not suffer from overfitting and can across multiple bridges. 
We verify the DS~\&~DI properties of our feature by visualization and by comparing it to other time- and frequency-domain features for estimating stiffness reductions and locations.

The main contributions of this paper are 1) We cast the IBHM problem as a signal decomposition, thus affording us the tools from the signal processing community; 2) We use the synchrosqueezed wavelet transform to extract a DS~\&~DI feature for IBHM; and 3) We validate the DS~\&~DI properties of it through extensive experiments.

\vspace{-3mm}
\section{Physical foundations}
\vspace{-2mm}
To model the VBIS, we consider a commonly used model, a sprung mass (representing vehicle) traveling with a constant speed on a simply supported beam (representing bridge). This model provides physical foundations of the IBHM problem, which help us to formulate this problem as a signal-decomposition problem.

\mypar{Vehicle-bridge interaction system}
The derivation of the theoretical formulation of the VBIS follows the same assumptions and system geometry as presented in \cite{liu2019damage}, albeit a local damage is considered in this work.

Let $x$ be the coordinate of the beam with the origin at the left support; $\rho,~A,~\mu,~EI(x)$ the density, cross section area, damping coefficient and stiffness of the beam, respectively; $\delta(x)$ the Dirac delta function; $g$ the gravity constant; $v$ the moving speed of the vehicle; and $y(t)$ the vertical displacement of the vehicle chassis; $m_v,~k_v,~c_v$ the weight, stiffness and damping coefficient of the vehicle, respectively. The vertical acceleration of the vehicle chassis, $\ddot{y}(t)$ is our measurements. The equations of motion for the VBIS are
\vspace{-0.5em}\begin{equation}
    \label{eq_bmotion}
\begin{alignedat}{2}
    \rho A\frac{\partial^2 u(x,t)}{\partial t^2}&+\mu\frac{\partial u(x,t)}{\partial t}+\frac{\partial^2}{\partial x^2}\Big(EI(x)\frac{\partial^2 u(x,t)}{\partial x^2}\Big)\\
    &=\delta(x-vt)(m_vg+m_v\ddot{y}(t)),
\end{alignedat}
\end{equation}
\vspace{-0.5em}
\begin{equation}
\label{eq_vmotion}
\begin{alignedat}{2}
    m_v\ddot{y}(t)+&c_v\dot{y}(t)+k_vy(t)=c_v\dot{u}(vt,t)+k_vu(vt,t).
    \end{alignedat}\vspace{-0.3em}
\end{equation}
Using modal superposition, we write
$u(x,t)=\sum_{n=1}^{\infty}\phi_n(x)q_n(t)$,
where $\phi_n(x),~q_n(t)$ are the $n$-th mode shape and mode displacement, respectively.

The stiffness of the beam with local damage is defined as\vspace{-0.5em}
\begin{equation}
\label{eq_stiffness}
    EI(x)=\begin{cases}
    EI_0 & \text{if}\quad x\not\in [x_s-\frac{l_s}{2},x_s+\frac{l_s}{2}]\\
    EI_0(1-R_s) & \text{if}\quad x\in [x_s-\frac{l_s}{2},x_s+\frac{l_s}{2}]
    \end{cases}\vspace{-0.5em}
\end{equation}
where $x_s$ is the central location of the damage; $l_s$ and $R_{s}$ are the damage length and the percentage reduction of the stiffness, respectively; $EI_0$ is the stiffness of the undamaged beam segment. $x_s$ and $R_s$ are the parameters we want to infer.
Note that many previous works~\cite{yang2005vehicle,yang2014constructing,yang2015extraction,liu2019damage} consider the stiffness as a constant. In this paper, we solve the VBIS with the stiffness-reduction-type local damage.

\mypar{Damage-sensitive and domain-invariant features}
\label{sec:signal}
We now solve Eq.~\eqref{eq_vmotion}, look for a DS~\&~DI feature and formulate the IBHM problem as a signal decomposition problem.  By omitting the damping of the beam and the vehicle, we write the acceleration of the moving vehicle as\vspace{-0.5em}
\begin{equation}
\label{veh_solution}
\begin{alignedat}{2}
    \ddot{y}(t) =&\sum_{n=1}^{\infty}C_{1n}\sin(\omega_v t)+\sum_{n=1}^{\infty}\big[C_{2n}\phi_n(vt)\sin(\tilde{\omega}_nt)\\
    &+C_{3n}\dot{\phi}_n(vt)\cos(\tilde{\omega}_nt)+C_{4n}\ddot{\phi}_n(vt)\sin(\tilde{\omega}_nt)\big]\\
    &+\sum_{n=1}^{\infty}C_{5n}(\phi_n(vt)\ddot{\phi}_n(vt)+\dot{\phi}_n(vt)^2),
\end{alignedat}\vspace{-0.5em}
\end{equation}
where $C_{1n}$, $C_{2n}$, $C_{3n}$, $C_{4n}$, and $C_{5n}$ are constants depending on properties of the bridge and the vehicle;  $\tilde{\omega}_n=2\pi f_n=\sqrt{\tilde{k}_n/\tilde{m}_n}$, $\tilde{m}_n=\int_0^L\rho A\phi_n^2(x) d x$,  $\tilde{k}_n=\int_0^LEI(x)(\phi_n '' (x))^2 d x$
are the $n$-th mode's equivalent resonant frequency, mass, and stiffness, respectively; and $\omega_v=2\pi f_v$ is the natural frequency of the vehicle. Note that when we solve Eq.~\eqref{eq_vmotion}, because $m_v<<\tilde{m}_n$, we have 
$\frac{m_vg+m_v\ddot{y}(t)}{\tilde{m}_n}\approx \frac{m_vg}{\tilde{m}_n}$.

For our VBIS, local damage has relatively small influence on the bridge mode shape. Thus, we can approximate $C_{51}\appropto \frac{\omega_v^2\tilde{\omega}_1^2m_vg}{\tilde{k}_1(\tilde{\omega}_1^2-\omega_{d1}^2)}\frac{\omega_{d2}^2}{(\omega_v^2-\omega_{d2}^2)}$, where $\omega_{dn}=2\pi f_{dn}=n\pi v/L$.

We can obtain a physical understanding of the vehicle acceleration by analyzing Eq.~\eqref{veh_solution}. First, Eq.~\eqref{veh_solution} has the following three components within three frequency bands:
\vspace{-0.2em}
\begin{itemize}
\itemsep0em 
    \item $\sum_{n=1}^{\infty}C_{1n}\sin(\omega_v t)$ may encoder damage information in $C_{1n}$. The dominant frequency of this term is $\omega_v$, which changes as vehicle properties change;
    \item $\sum_{n=1}^{\infty}\big[C_{2n}\phi_n(vt)\sin(\tilde{\omega}_nt)+C_{3n}\dot{\phi}_n\cos(\tilde{\omega}_nt)+C_{4n}\ddot{\phi}_n(t)\sin(\tilde{\omega}_nt)\big]$ may encode damage information in $C_{2n}$, $C_{3n}$, $C_{4n}$, $\tilde{\omega}_n$ and $\omega_{\beta n}(t)$. The dominant frequency of this term is $\tilde{\omega}_n\pm \omega_{\beta n}(t)$, which changes as bridge properties change;
    \item $\sum_{n=1}^{\infty}C_{5n}(\phi_n(vt)\ddot{\phi}_n(vt)+\dot{\phi}_n(vt)^2)$ may encode damage information in both $C_{5n}$ and $\omega_{\beta n}(t)$, and the multiplication of the derivatives of $\phi_n(vt)$ amplifies the damage information. Also, the dominant frequency of this term is $2\omega_{\beta n}(t)$, which does not change as the bridge and vehicle properties change.
\end{itemize}
\vspace{-0.2em}
For each $n$, the third component is thus DS~\&~DI once we multiply it by factor $1/C_{5n}$. Second, because of the local influence of the damage, $\phi_n(vt)$ is non-stationary and its instantaneous frequency, which is defined as $\omega_{\beta n}(t)$, is time-varying and $\geq \omega_{dn}$. Third, for VBIS, $\omega_{\beta n}(t)$ is very small and generally does not overlap with the vehicle frequency $\omega_v$ and the shifting bridge frequency $\tilde{\omega}_n\pm\omega_{bn}(t)$. This property of the vehicle acceleration signal also indicates that the third component is a good candidate for our desired feature because it can be extracted from the original signal.

\vspace{-3mm}
\section{Signal decomposition method for IBHM}
\vspace{-2mm}
To extract the DS~\&~DI feature from the vehicle acceleration signal, we can cast the IBHM problem as a signal decomposition problem. We decompose the non-stationary vehicle acceleration and extract the desired feature by reconstructing the non-stationary component within a frequency band that includes $2\omega_{\beta n}(t)$. We consider approximating  the DS~\&~DI feature for the first mode that is $
    y_d(t)=\phi_1(vt)\ddot{\phi}_1(vt)+\dot{\phi}_1(vt)^2.$
Note $\phi_1(vt)$ has time-varying instantaneous frequencies. Also, $y_d(t)$ is the sum of harmonic functions, hyperbolic functions and multiplication of harmonic and hyperbolic functions, so that the desired feature can be expressed as a superposition of IMFs. 

Our proposed method uses SWT~\cite{daubechies2009synchrosqueezed,daubechies2011synchrosqueezed} to represent the vehicle acceleration signal in the time-frequency plane and reconstruct the desired feature using ISWT. The reconstructed feature is used to estimate and localize damage. SWT has three advantages in solving our problem. First, it assumes that signals are approximately harmonic locally and have a slowly time-varying instantaneous frequency. This transform has the ability to decompose a non-stationary and time-varying signal
as a superposition of IMF-type components. Second, comparing with the conventional time-frequency methods, such as short-time Fourier transform (STFT) and continuous wavelet transform (CWT), this empirical model decomposition (EMD)-like approach can further sharpen the time-frequency representation and enhance frequency localization~\cite{daubechies2011synchrosqueezed}. Third, to localize the damage, we need to reconstruct the time-domain signal in the damage-related frequency band as the vehicle moves. ISWT can directly reconstruct a component within a selected frequency band and avoid the mode-mixing effect encountered by the EMD method~\cite{mandic2013empirical}. The SWT has three steps: 

1) Calculate the wavelet coefficients of the signal
\vspace{-0.6em}\begin{equation}
\begin{alignedat}{2}W_y(a,b)=\frac{1}{\sqrt{a}}\int_{-\infty}^{\infty}\ddot{y}(t)\Phi^*\big(\frac{t-b}{a}\big)dt,\end{alignedat}\end{equation}\vspace{-0.2em}
where $a$ is the scale, $b$ is the time offset, and $\Phi^*(t)$ is the complex conjugate of wavelet. We use analytic Morlet wavelet;

2) Estimate the instantaneous frequencies for the signal\vspace{-0.5em}
\begin{equation}
\begin{alignedat}{2}\omega_y(a,b)=
\frac{-j\partial_b [W_y(a,b)]}{W_y(a,b)},\qquad \text{for}~ W_y(a,b)\neq 0;\vspace{-0.5em}\end{alignedat}\end{equation}

3) Reallocate the energy of CWT coefficients to enhance frequency localization. At discrete scales $a_k$, the SWT of $\ddot{y}(t)$ is only calculated at the centers $\omega_c$ of the frequency range $\omega_c\pm \Delta\omega_c$, where $\Delta\omega_c=\omega_c-\omega_{c-1}$. The SWT is
\vspace{-0.5em}\begin{equation}
\begin{alignedat}{2}T_y(\omega_c,b)=\frac{1}{\Delta\omega_c}\sum_{a_k:|\omega_y(a_k,b)-\omega_c|\leq \Delta\omega_c/2}W_y(a_k,b)a_k^{-3/2}\Delta a_k,\vspace{-0.5em}\end{alignedat}\end{equation}
where $\Delta a_k=a_k-a_{k-1}$.  Component of the original signal in band $[\omega_{\beta 1}^l,\omega_{\beta 1}^u]$ is estimated by ISWT:\vspace{-0.5em}
\begin{equation}
\begin{alignedat}{2}\hat{\ddot{y}}(t)=\mathcal{R}\Big[1/2\int_0^{\infty}\hat{\Phi}^*(\xi)\frac{d\xi}{\xi}\sum_{\omega_c\in [\omega_{\beta 1}^l,\omega_{\beta 1}^u]} T_y(\omega_c,t)\Big],\vspace{-0.5em}\end{alignedat}\end{equation}
where $\mathcal{R}[\cdot]$ returns the real part of the function; $\hat{\Phi}^*(\xi)$ represents the Fourier transform of the complex conjugate of the wavelet. Further details about the implementation of SWT and ISWT can be found in~\cite{daubechies2009synchrosqueezed,daubechies2011synchrosqueezed}.
Our feature extraction algorithm is shown in Algorithm~\ref{at:1}.

\vspace{-0.6em}
\begin{algorithm}[ht]
\caption{DS~\&~DI feature extraction for IBHM}
\label{at:1}
\begin{algorithmic}[1]
\Require{Initialize known parameters: $m_v$, $k_v$, $v$, $L$, $\omega_{dn}$}
\State{\textbf{Input}: vertical acceleration of the moving vehicle: ${\ddot{y}}(t)$}
\State{Estimate system properties, including $\tilde{\omega}_1$ and $\tilde{k}_1$, using system identification methods, e.g. \cite{matarazzo2018scalable,liu2019expectation};}
\State{Compute the SWT, $T_y(\omega_c,b)$, for vehicle acceleration;}
\State{Choose a frequency band, $[\omega_{\beta 1}^l,\omega_{\beta 1}^u]$, for extracting the desired feature based on the identified $\omega_{d1}$;}
\State{Calculate ISWT, $\hat{\ddot{y}}(t)$,
  within the selected frequency band in order to approximately reconstructing the third component of Eq. \ref{veh_solution};}
\State{\textbf{Output}: the DS~\&~DI feature $y_d(t)=\hat{\ddot{y}}(t)/C_{51}$}
\end{algorithmic}
\end{algorithm}

\vspace{-3mm}
\section{Experimental results}
\label{sec:exp}
\vspace{-2mm}
\mypar{Finite element simulations and the dataset}
Finite element models (FEMs) of the VBIS are employed to create a dataset. We have five bridges (Bridge 1, 2, 3, 4 \& 5) simulated as Euler-Bernoulli beams of lengths $25 ~m$, $25 ~m$, $25 ~m$, $20 ~m$ and $30 ~m$, respectively; natural frequency of $2n^2 ~Hz$, $2.5n^2 ~Hz$, $3 n^2 ~Hz$, $2.5n^2 ~Hz$ and $2.5n^2 ~Hz$ (for $n= 1\cdots 10$), respectively; and they all have the same undamaged stiffness, $EI_0=14.54~GPa\times m^4$. We use the same properties (with reference to~\cite{yang2015extraction}) for the simulated vehicle: $m_v=100~kg$, $f_v=6.5~Hz$, and $v=3~m/s$.



Damage proxy is introduced by reducing the stiffness of one element at different locations. We use the same beam element with the length of $0.6~m$ for all bridges. The stiffness reduction ($R_s$) ranges from $70\%$ to $30\%$ with an interval of $10\%$. For each stiffness reduction level, the simulation is run at seven different damage locations ($x_s$ is every eighth of the span), and for each damage scenario, the simulation is run ten times. For each simulation run, random forces, which follows a Gaussian distribution with zero mean and $0.1~N$ variance, are applied on each node of the FEMs. This added disturbing force is a process noise that propagates through time and varies in space. In total, we generate $5$ (bridges) $\times$ $5$ (reduction levels) $\times$ $[7$ (damage locations)$+1$ (undamaged bridge)] $\times$ $10$ (trials) $=2000$ vehicle acceleration records.

\begin{figure}[htb]
\begin{minipage}[b]{.48\linewidth}
  \centering
  \centerline{\includegraphics[height=2.2cm]{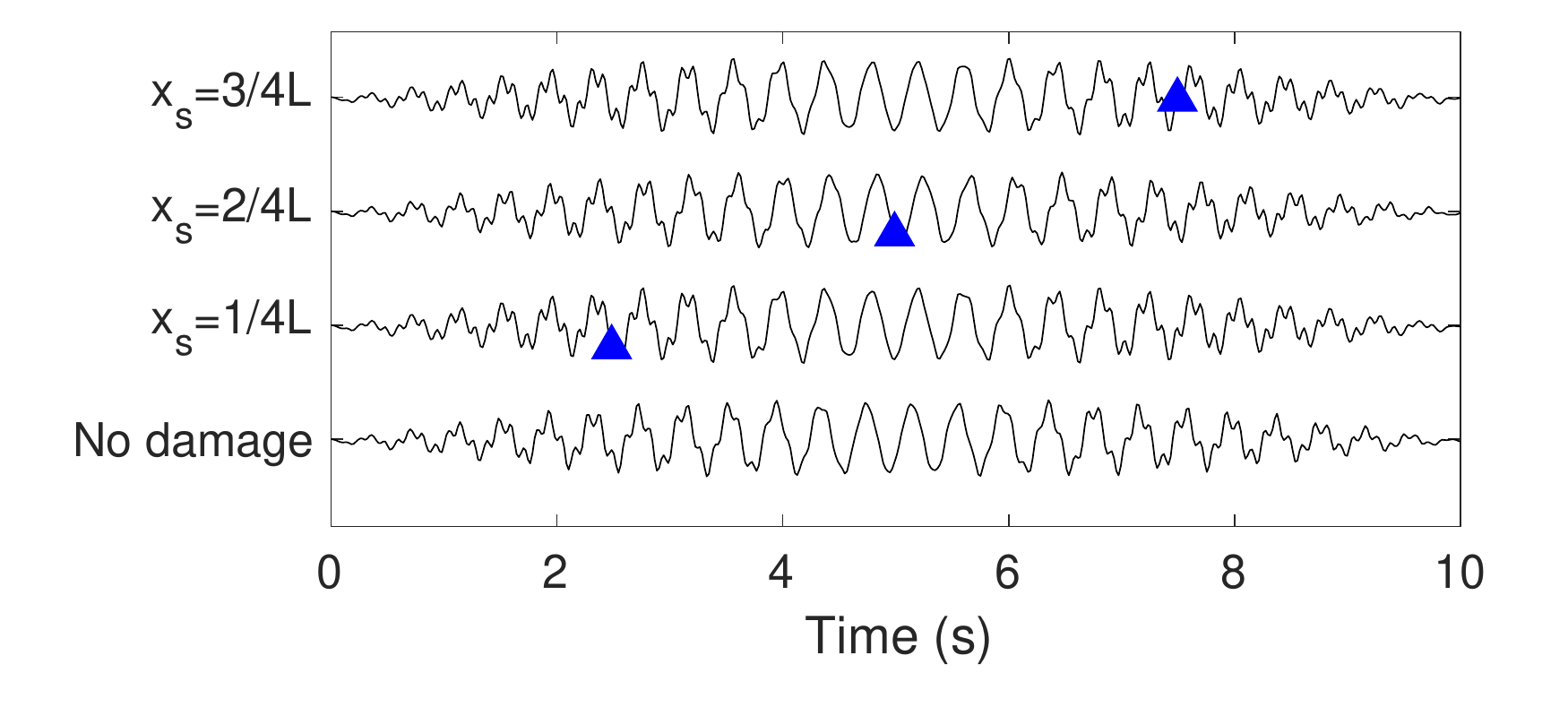}}
  \centerline{(a) Vehicle accelerations}\medskip
\end{minipage}
\hfill
\begin{minipage}[b]{0.48\linewidth}
  \centering
  \centerline{\includegraphics[height=2.2cm]{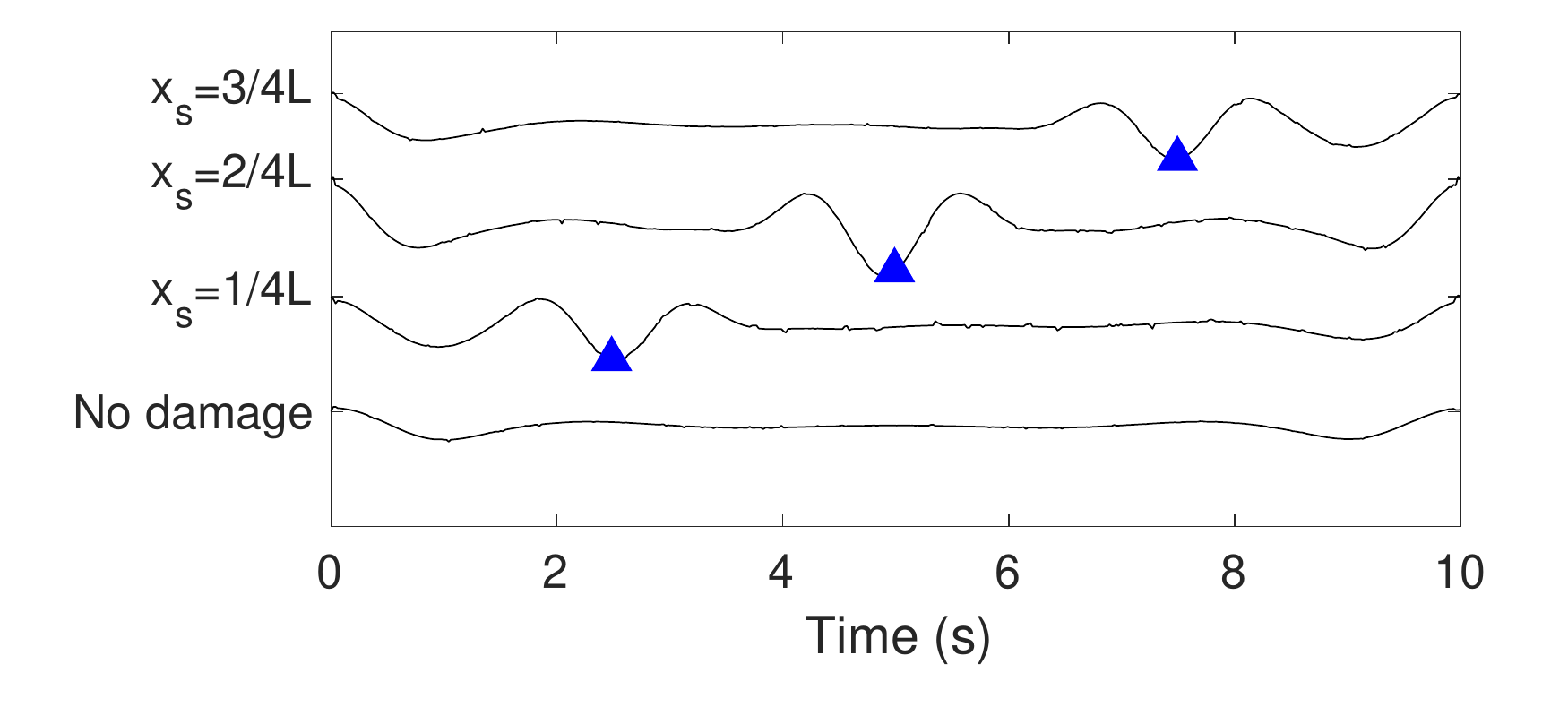}}
  \centerline{(c) Our proposed feature}\medskip
\end{minipage}
\vfill
  \vspace{-.6em}
\begin{minipage}[b]{.48\linewidth}
  \centering
  \centerline{\includegraphics[height=2.2cm]{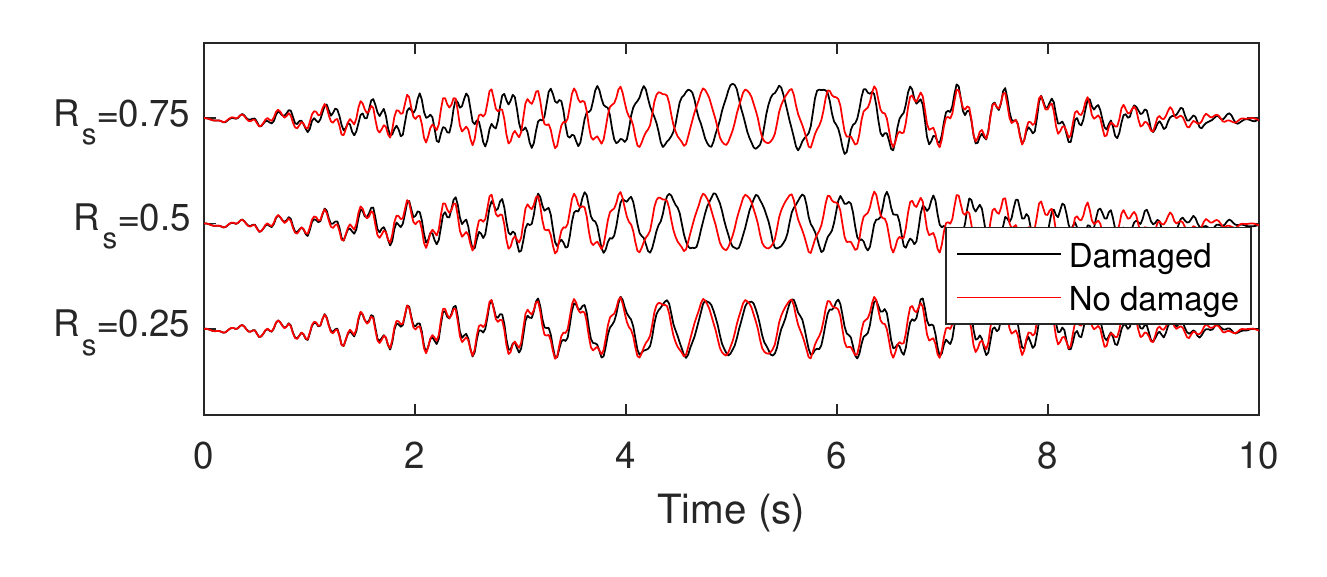}}
  \centerline{(c) Vehicle accelerations}\medskip
\end{minipage}
\hfill
\begin{minipage}[b]{0.48\linewidth}
  \centering
  \centerline{\includegraphics[height=2.2cm]{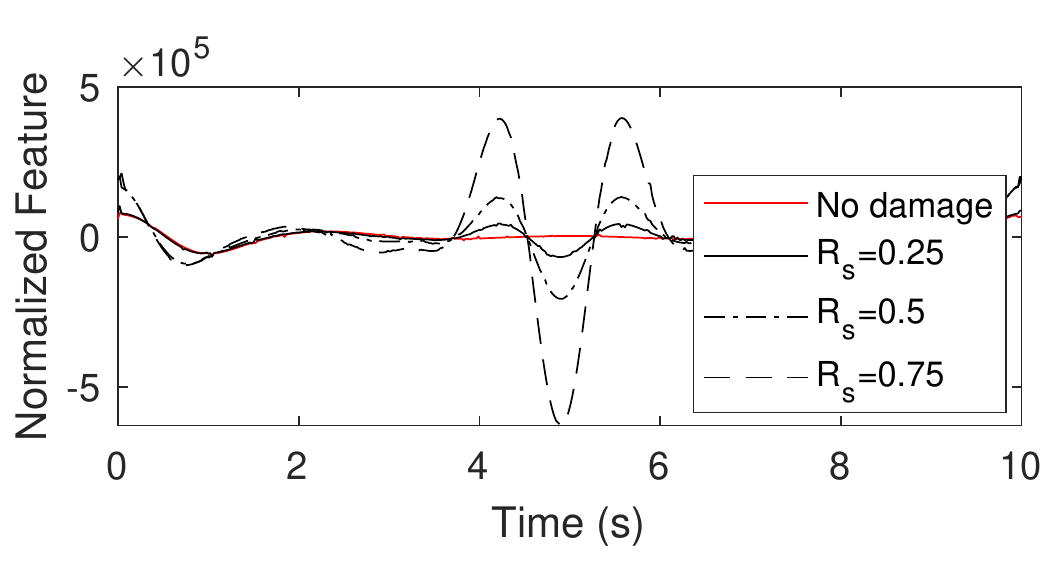}}
  \centerline{(d) Our proposed feature}\medskip
\end{minipage}
  \vspace{-1.em}
\caption{Raw signals and our proposed feature for vehicle accelerations with different damage locations (a, and b) and different stiffness reduction levels (c and d). The blue marks indicate the damages. We can visually localize and compare the stiffness reductions using our proposed feature, which verifies that our feature is DS.}
\label{fig:verify}
  \vspace{-0.5em}
\end{figure}

\begin{figure}[htb]
\begin{minipage}[b]{0.48\linewidth}
  \centering
  \centerline{\includegraphics[height=1.9cm]{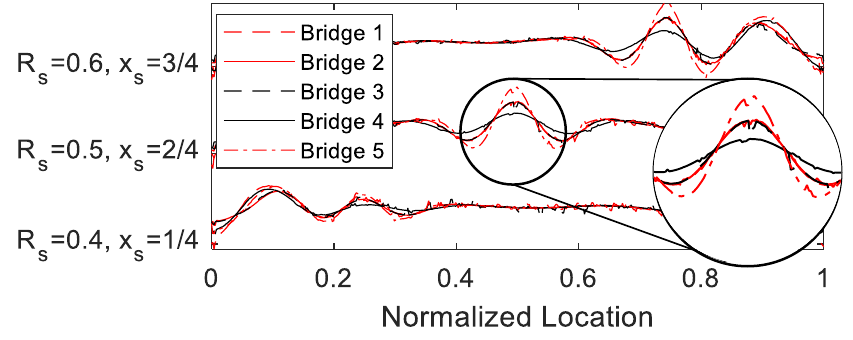}}
  \centerline{(a) ISWT in $[f_{d1},1 Hz]$}\medskip
\end{minipage}
\hfill
\begin{minipage}[b]{0.48\linewidth}
  \centering
  \centerline{\includegraphics[height=1.9cm]{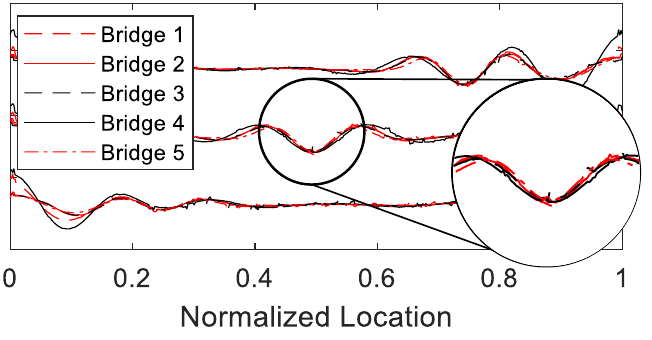}}
  \centerline{(b) Our proposed feature}\medskip
\end{minipage}
  \vspace{-1.em}
\caption{ISWT within $[f_{d1},1 Hz]$ of vehicle accelerations traveling on Bridges 1, 2, 3, 4 \& 5 with different damage (a) before and (b) after multiplying by $1/C_{51}$. Our features for bridges with different properties and the same damage match each other around damage locations. However, the features before applying the proposed multiplication do not match each other. This visualization verifies that our feature is DI.}
\label{fig:verify1}
  \vspace{-0.5em}
\end{figure}

\mypar{Visualization of the proposed feature}
We verify the DS~\&~DI properties of our feature by visualization. 
The data for visualization is created by FEMs without adding the disturbing force. In our experiments, we use the proposed method to extract the desired feature within $[f_{d1}, 1~Hz]$. Figures~\ref{fig:verify} (a) and (b) show vehicle accelerations and our proposed feature with different damage locations, respectively. Figures~\ref{fig:verify} (c) and (d) show the signals and the feature with different stiffness reductions. By visualization, we can easily localize and compare the damage simulated by reducing stiffness. To verify if the proposed feature is DI, in Figure~\ref{fig:verify1} we visualize the ISWT of accelerations collected from the simulated vehicle traveling on Bridges 1 to 5 with different simulated damage before and after multiplying by $1/C_{51}$. We can observe that, after the multiplication, the features for different bridges having the same damage match each other, which indicates that our feature is DI. Though the results also contain boundary effects of the transform that require further investigation, it is easy to see that at the normalized locations of damage the proposed feature exhibit high sensitivity.

\vspace{-0.5em}

\mypar{Damage localization and quantification} To further verify if our proposed feature is DS~\&~DI, we use a multi-task learning model proposed in~\cite{liu2019damage} to estimate and localize stiffness reductions in a supervised and a semi-supervised fashion. 
The model's input includes raw accelerations (raw data); band-pass filtered signals within $[f_{d1},1 ~Hz]$, $f_n\pm 0.5 ~Hz$, and $f_v\pm 0.5 ~Hz$ (Bandpass 1, 2\&3); inverse CWT within the above three bands (ICWT 1, 2~\&~3); the first three IMFs calculated using EMD (IMF 1, 2~\&~3); ISWT within the three bands (Ours, ISWT 2~\&~3); and spectrograms calculated using STFT, CWT and SWT.

\begin{table}[ht]
\centering
  \vspace{-1em}
\caption{Damage localization and quantification results in terms of RMSE. Lower error means better performance.}
  \vspace{-.5em}
\label{results}
\begin{tabular}[t]{lccccccc}
\toprule
\multirow{2}{*}{Feature}&\multicolumn{2}{c}{Supervised}&\multicolumn{2}{c}{Different $L$}&\multicolumn{2}{c}{Different $\tilde{\omega}_1$}\\
\cline{2-3}\cline{4-5}\cline{6-7}
& DLE& SRE& DLE& SRE& DLE& SRE\\
\midrule
Raw data&0.28&0.15&0.70&0.51&0.57&0.55\\
Bandpass 1&0.32&0.26&0.56&{0.44}&0.48&0.49\\
Bandpass 2&0.26&0.16&0.39&0.42&0.67&0.69\\
Bandpass 3&0.38&0.31&0.46&0.41&0.64&0.91\\
STFT &0.59&0.15&0.71&0.34&0.47&0.35\\
\midrule
ICWT 1&0.18&\textbf{0.08}&0.37&\textbf{0.27}&0.38&{0.28}\\
ICWT 2&0.24&0.22&0.40&0.35&0.52&0.37\\
ICWT 3&0.26&0.17&0.56&0.51&0.37&0.34\\
CWT&0.43&0.52&0.48&0.61&0.63&0.31\\
\midrule
IMF 1&0.28&0.21&\textbf{0.36}&0.34&\textbf{0.33}&0.36\\
IMF 2&0.27&0.45&0.40&0.56&0.82&0.84\\
IMF 3&0.35&0.28&0.52&0.43&0.44&0.28\\
\midrule
\textit{\textbf{Ours}}&{\textbf{0.17}}&{\textbf{0.08}}&\textbf{0.36}&{\textbf{0.27}}&{0.38}&\textbf{0.27}\\
ISWT 2&0.30&0.26&0.39&0.39&0.55&0.53\\
ISWT 3&0.28&0.18&0.39&0.33&0.58&0.30\\
SWT&0.56&0.38&0.62&0.45&0.43&0.65\\
\bottomrule
\end{tabular}
\vspace{0.2ex}

{\raggedright \small 1. DLE means damage location estimation.\\
2. SRE means stiffnens reduction estimation\par}
\vspace{-2mm}
\end{table}%

The supervised task examines how sensitive each input is to damage. The second and third columns in Table~\ref{results} present this task's results (30\% for testing) in terms of root mean squared error (RMSE). Using the proposed feature, we obtain the best stiffness reduction estimation and localization results. 

Table~\ref{results} also shows results for the semi-supervised regressions, where we have two sub-tasks: test if the feature is DI across bridges with different lengths (Different $L$) and with different natural frequencies (Different $\tilde{\omega}_1$). As shown in columns 4 to 7 of the table,
using our feature, we obtain the best stiffness reduction estimations for the two sub-tasks, and the best damage localization results for bridges having different lengths and the same frequency. For damage localization in the second sub-task, IMF 1 provides the best result.

\vspace{-3mm}
\section{Conclusion}
\label{sec:con}
\vspace{-2mm}
We introduce a physics-guided signal decomposition method to extract a DS~\&~DI feature from vehicle accelerations for IBHM. The SWT is used to represent the data in the time-frequency plane, and the desired feature is reconstructed using ISWT within a damage-related frequency band. We verify and evaluate the DS~\&~DI properties of the extracted feature for IBHM using the simulated data generated from FEMs. Among the six experiments we conducted, five of them exhibited the best damage quantification and localization results across different bridges using our proposed feature.

\vfill
\pagebreak
\bibliographystyle{IEEEbib}
\bibliography{refs}

\end{document}